\documentclass{aa}
\usepackage{natbib}
\usepackage{graphicx}
\usepackage{txfonts}
\def\arcsec{\hbox{$^{\prime\prime}$}}

\def\degree{^\circ}

\begin{document}
   \title{Scattered Lyman-$\alpha$ radiation of comet 2012/S1 (ISON) observed by SUMER/SOHO}
   \author{W. Curdt$^{1}$, H. Boehnhardt$^{1}$, J.-B. Vincent$^{1}$, S.K. Solanki$^{1}$,
   U. Sch\"uhle$^{1}$, L. Teriaca$^{1}$
       }

\institute{$^{1}$ Max Planck Institute for Solar System Research, Justus-von-Liebig-Weg 3, 37077 G\"ottingen, Germany\\
                     email: {\sf curdt@mps.mpg.de}\\
             }

\authorrunning{Curdt et al.}
\titlerunning{SUMER Lyman-$\alpha$ observations of comet ISON}
%\email{curdt@mps.mpg.de}
%}

%   \institute{Max-Planck-Institut f\"ur Sonnensystemforschung,
%   Max-Planck-Str. 2, 37191 Katlenburg-Lindau, Germany\\

%  \email{curdt@mps.mpg.de}

   \date{Received May 8, 2014; accepted May 9, 2014}
\abstract
   {During its sungrazing perihelion passage, comet ISON
   appeared in the field of view of the SUMER spectrometer and allowed
   unique observations at far-ultraviolet wavelengths with high spatial and temporal resolution.
   We report results of these observations completed on November 28, 2013, when the
   comet was only 2.82~$R_{\scriptscriptstyle \bigodot}$ away from the Sun.
   Our data show the arrow-shaped dust tail in Ly-$\alpha$ emission trailing
   behind the predicted position of the nucleus, but offset from the trajectory.
   We interpret the emission as sunlight that is scattered at $\mu$m-sized dust particles.
   We modeled the dust emission and dynamics to reproduce the appearance of the tail.
   We were unable to detect any signature of cometary gas or plasma around the expected position
   of the nucleus and conclude that the outgassing processes must have stopped
   before the observation started. Moreover, the model we used to reproduce the
   observed dust tail needs a sharp fall-off of the dust production hours before perihelion transit.
   We compare the radiances of the disk and the dust tail for an estimate of the
   dust column density and tail mass.
   %After observing 18~years mostly solar targets, this was the first time that SUMER completed comet observations.
   }
   {}
%
 %  \keywords{Sun: corona,
  %           comet}
   \maketitle
%
%________________________________________________________________
\section{Introduction}

Comet 2012/S1 (ISON) -- believed to originate from the Oort Cloud and approaching
the Sun in a sungrazing orbit -- is a very special case that excited the
community, but also left it with a bundle of question marks. The lack of
comparable events in the past spawned non-converging predictions and
wild speculations. It was clear that ISON, coming as close as 2.7~$R_{\scriptscriptstyle \bigodot}$
to the Sun, would offer a unique chance for cometary science and in particular
for spectroscopy. Oort-Cloud comets are ideal objects to study compositional information
and outgassing processes, and a wordwide campaign was initiated to observe the comet
in various bands of the electromagnetic spectrum (e.g., Knight et al.,
2014; Druckm\"uller et al., 2014, Ag\'undez et al., 2014).
Yet, ground-based night-time telescopes and instruments in
near-Earth-orbit often have limitations when they are pointed close to the Sun.
ISON, however, came so close that solar telescopes were able to close that gap.

\section{Observations: images and spectra}

Even early orbit predictions showed that ISON's trajectory
would cross the field-of-view (FOV) of the Solar Ultraviolet Measurements of Emitted
Radiation (SUMER; Wilhelm et al. 1995) spectrometer on SOHO
during perihelion transit, offering a unique chance of high-resolution
spectroscopic observations at far-ultraviolet (FUV) wavelengths.
We transformed the latest ephemerides (JPL orbital elements set \#54) to
SUMER coordinates using the attitude and orbit information of SOHO. Figure~1 depicts the
comet trajectory across the SUMER FOV in SOHO coordinates from a start position
$x$=-894\arcsec, {$y$=-1768\arcsec} to a final position $x$=-1644\arcsec, {$y$=701\arcsec}
imposed by the mechanism limits.
Each dot in Fig.~1 stands for the predicted position of the central brightness at
a given minute. The comet was expected to enter the FOV at 17:56~UTC.

%figure 1
\begin{figure}
   \centering
   \includegraphics[width=5cm]{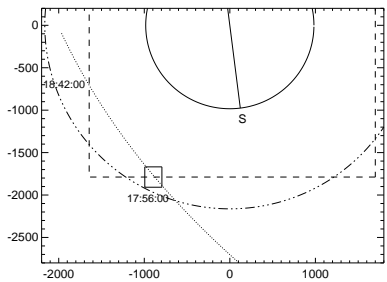}
      \caption{Trajectory of comet ISON across the SUMER FOV in the SOHO coordinate system (units of arcseconds).
      SOHO's attitude is
      oriented toward ecliptic north. For each minute the predicted position of the central brightness
      is marked by a dot. The accessible FOV of SUMER -- here indicated by a
      dashed line -- refers to the central pixel
      of the slit. The instantaneously imaged area of 200\arcsec $\times$ {240\arcsec}
      is just inside the LASCO-C2 occulter (dash-dotted circle).
      }

\label{orbit}
\end{figure}

We completed limb tests to improve the pointing accuracy and assumed that the uncertainly
is {$\approx$5\arcsec} in azimuth and {$\approx$10\arcsec} in elevation. The
uncertainty of the orbit prediction is even smaller (Giorgini 2014, private communication).

{\bf Images:} first, SUMER was operated in slot mode, that is, with a 2~mm hole instead of the slit.
In this mode, the telescope provides 2D stigmatic images that are in $x$-direction convolved with the Ly-$\alpha$
line profile. From each exposure of 15~s a window of 200 px $\times$ 240 px was
flatfield-corrected on board and downlinked in real time with a 2~$\times$~2 pixel binning.
%The pointing information in Fig.~1 refers to the central pixel of the image.
%Therefore, the bottom pixels extend the range of the FOV towards the South.

Altogether, 101 images were obtained in this mode. The first 70 images show no
cometary signal. We calculated an average background image from these images
for straylight subtraction in the remaining images. Finally, a median~10 filter
was applied for noise reduction. In Fig.~2 a series of 12 images -- each one the sum
of two exposures -- shows the entry of the comet into the SUMER FOV. The
orbit was almost tangential at that time, and distance to perihelion was only
0.15~$R_{\scriptscriptstyle \bigodot}$.
A cross in each image marks the predicted position of the nucleus. The granulation
along the feature is most likely an artifact, a combination of residual detector
inhomogeneity and noise-filtering. The noise in the bottom pixels stems from an
edge problem of the noise filter.

%figure 2
\begin{figure}
   \centering
   \includegraphics[width=9cm]{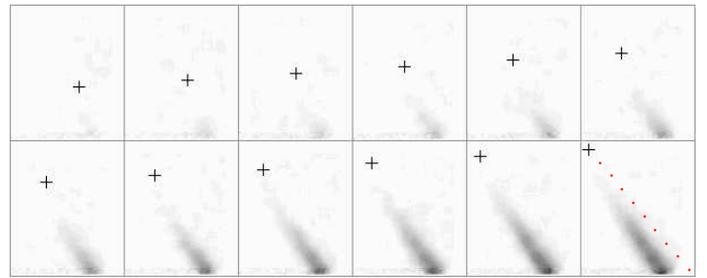}
      \caption{Dust tail of comet ISON around perihelion transit observed in Ly-$\alpha$.
      These 12 slot images with a size of 200\arcsec $\times$ 240\arcsec were taken
      with a cadence of 30 s between 17:56 and 18:01 UTC (top left to bottom right).
      The cross marks the predicted position of the nucleus.
      In the final image red dots indicate the trajectory in steps of 30~s.}
\label{perihel}
\end{figure}

{\bf Spectra:} at 18:02 UTC the slit-less sequence was stopped and the spectroscopic mode started,
employing the {4\arcsec} slit. The prime goal of this mode was the spectroscopic
analysis of gas and plasma emission in the coma and gaining compositional information.
Our sequence cycled through four spectral windows around 105~nm, 110~nm, 121~nm, and 131~nm,
chosen for emission lines of molecules, atoms, and ions such as O\,{\sc vi}.
We repointed every ten minutes to track the comet. However, most of these spectra had an extremely
low signal, only few spectra showed signal of scattered light from the solar disk.
We can exclude that emission of cometary gas or plasma above detection limit
was recorded then. The lack of any signature in SDO-AIA images
corroborates this finding.

\subsection{Comet morphology}

The very short ionization time of neutral hydrogen so close to the Sun sets an
upper limit to the extent of the neutral hydrogen coma in Ly-$\alpha$ at a few 100~km, well
within one SUMER pixel. Moreover,
the number of hydrogen atoms in the \mbox{Ly-$\alpha$} emission state is also reduced
by orders of magnitude (compared to a similar case farther away from  the Sun) as our Haser calculations have shown.
Such considerations also apply for other gaseous species, thus excluding
emission from genuine neutrals as well as from
charge-exchange neutrals (Bemporad et al. 2005).
Hence, the measured radiance is attributed to sunlight that is reflected and
scattered by the cometary dust, since thermal emission of the dust at FUV wavelengths
is negligible.

%figure 3
\begin{figure}
   \centering
   \includegraphics[width=9cm]{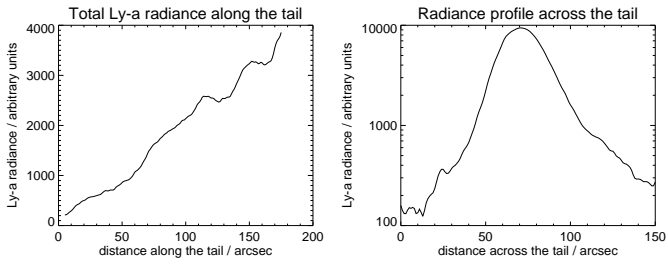}
      \caption{
      Radiance profile of the coma along and across the tail.}
\label{profiles}
\end{figure}

The Ly-$\alpha$ emission above background level shows a spiky diffuse arrow-shaped
coma with a tip in ram direction. No local brightness condensation or any other
indication of the nucleus is found above detection limit. The spiky coma lags behind the predicted
position. Pointing uncertainty and orbit prediction uncertainty cannot explain
this discrepancy. The coma and tail region is offset from the expected trajectory
in antisolar direction and appears at a small angle to the comet path.
The brightness increase along the feature is approximately linear
from the tip to maximum level close to the edge of the image (see Figs.~3 and 4a).
The slope of the brightness increase is about 1.7 relative units per 100~Mm.
The elongated coma extends over at least 240~Mm with a possible peak brightness
at or beyond the lower edge of the image.
The core isophotes of the coma spike are symmetric with respect to the main axis of the coma pattern.
This main axis is rather straight and is oriented along a position angle $PA$ of $72\degree$
with respect to celestial North. The lower level isophotes are
asymmetric with respect to this line and appear to be more extended in the direction of the Sun.
The spiky coma has an opening angle of about $13\degree \pm 1\degree$ as seen from the front end tip
and reaches a maximum width of more than 100~Mm.
Figure~4a depicts an enlarged copy of the last image in Fig.~2. Again, the
cross marks the predicted position of the nucleus at mid-exposure, and ten red dots mark its
position at one-minute intervals. It is obvious
that the tail is offset from the trajectory by {$\approx$30\arcsec} in a direction away
from the Sun. This is three times more than the pointing uncertainty. We also
studied the cross-section of the tail. At ten positions we determined the
geometrical center of brightness across the tail. In Fig.~4a these locations are marked by
white dots. The bi-sector -- the line connecting these dots -- is not aligned with the trajectory,
but appears at an angle of {$\approx$$5\degree$} in anti-solar direction from the trajectory.
It does not seem to be a perfect straight line either but has a curvature
toward the suspected nucleus position.

 \subsection{Radiometry}

Scattered light from the disk limits the detectability of off-disk targets
in unocculted solar telescopes. At the comet's position -- far
out at the edge of the SUMER FOV -- Ly-$\alpha$ radiation of the disk produces
a scattered-light level that is four orders of magnitude lower than
that of the average quiet Sun (Curdt et al. (2001).
The measured Ly-$\alpha$ background of 19.3~mW/(m$^2$ sr) is similar
to the radiance emerging from the coma, which is 18~mW/(m$^2$ sr) in the brighter sections.
Since we observed at a short wavelength of 121.6~nm, we assume that a geometric model
can be used for estimating the dust column density for $\mu$m-sized
particles. For an optically thin dust environment, the radiance of the dust tail, $L_{tail}$,  is given
by \

\vspace{2mm}$L_{tail}$ = $A$ $\eta$ $l$ $\pi r^{2}$ $L_{disk}$ ,\

\vspace{2mm} where $A$ denotes the phase-angle-dependent albedo, $\eta$ the
volumetric density of the particles, $l$ the column length,
$r$ the dust grain radius, and $L_{disk}$ the average radiance of the solar disk.
For simplification we assumed both of the latter to be uniform.
The Sun was extremely quiet during perihelion transit, its irradiance was only 11\% above
solar minimum levels\footnote{http://lasp.colorado.edu/lisird/lya/}.
Therefore, we can assume a solar radiance $L_{disk}$ of $\approx$101~W/(m$^2$
sr). With a column length of
{$\approx$10\arcsec} -- equivalent to 7.25~Mm -- and the tail being 1.8~10$^{-4}$
times fainter than the disk, the product $A$~$\eta$~$\pi r^{2}$ can be determined.
For an example case with a value of 0.1 for the albedo and a standard grain radius of 10~$\mu$m
a result of $\approx$0.78 particles~/~m$^3$ for $\eta$ is returned. This implies that in total
$\approx$2.7~10$^{21}$ standard grains are in the observed structure, equivalent
to a volume of 1.15~10$^7$ m$^3$ or a mass of 11.500 t of material with unit
density.

In the latest C2 images the shape of the comet is also pin-tipped
(cf., Fig.~3 in Druckm\"uller et al., 2014).
Despite this similarity, we were unable to reliably determine the same ratio for
visible-light to further constrain the grain size distribution. Coalignment
of images that are neither cospatial nor cotemporal is problematic, and
C2 data at altitudes $\le 3 R_{\scriptscriptstyle \bigodot}$ have significant
photometric uncertainties (Knight et al. 2010).

\section{Coma and tail modeling}

We employed dust dynamic models to reproduce the main features observed.
Finson-Probstein calculations (Finson \& Probstein 1968; Bei{\ss}er 1991) show
that the main axis of the coma spike with a position angle $PA$ of $72\degree$ follows a
synchrone with an emission time at 8.5~h $\pm$~0.5~h before perihelion transit.
The narrow opening angle of the coma spike ($\approx$$13\degree$) covers synchrones of
dust that might have been emitted within at most 12~h before perihelion transit.
We thus assume that the coma pattern seen in SUMER images resulted
from a short dust-emission event less than half a day before.
The dust production of the event decayed rapidly within a few hours.
As a consequence, one has to assume that the nucleus of the comet is located
close to the tip of the coma spike. The extension of the main spike axis
covers dust with a ratio $\beta$ for the solar radiation pressure to solar gravity
of up to 0.025, representative of basaltic dust of a few $\mu$m size and
larger, or graphite grains larger than 10~$\mu$m. We note that basaltic-type
grains of very small size ($<$0.02~$\mu$m) may have similar $\beta$ ratios,
but appear to be inefficient light scatterers at FUV wavelength.

The dust coma of a continuously active comet usually appears roundish with a typical
diameter of several 10~Mm and a pronounced brightness peak close to the center
that accommodates the nucleus. For comet ISON such a coma pattern with a diameter
of {$>$150\arcsec} and a sharp brightness falloff at its front was seen in
LASCO images until shortly before the extreme brightening on Nov 28.
This changed after the final brightening, the tail head narrowed continuously
and became almost pin-like,
comparable in width to a C2 pixel of 11.4\arcsec with no sharp edge at the tail
head anymore when it disappeared behind the C2 occulter.
The arrow-shaped spiky dust coma evolved immediately after the brightness maximum.
Brightness outbursts of several magnitudes normally indicate nucleus
fragmentation events (Boehnhardt 2004).
An extraordinary brightness peak with a maximum of -2 to -3mag
was measured in comet ISON on Nov~28 between 0 and 7~UTC (Knight \& Battams 2014).
Additional indications for extraordinary dust production around that time period
come from the preliminary analysis of the post-perihelion dust tail of comet
ISON (Boehnhardt 2013; Sekanina et al. 2013). It is very unlikely that the
observed spiky coma in SUMER and LASCO images is produced by a
chain of active nucleus fragments from the Nov~28 break-up.
Typical separation speeds from known nucleus splittings amount to less than 10~m/s,
which is insufficient for the extension of the coma spike in comet ISON (Boehnhardt 2004).
One would have to assume separation speeds of several km/s.

In a trial-and-error approach we used the Cosima dust simulation software
(Bei{\ss}er 1991) to model dust-tail images and isophote patterns in comet
ISON for the SUMER observing period with the goal to quantitatively describe the scenario outlined
above and to obtain further conclusions on the event and the released dust.
With a continuous dust production rate we were unable to reproduce anything
similar to what was observed.
The simulations show that an arrow-shaped coma with a position angle of
$72\degree$ for the main axis of the isophotes results for decreasing dust activity
after about 8h before perihelion transit (see Fig.~4b). The main axis is offset
by about 20~Mm to 30~Mm (projected into the sky) from the nucleus trajectory
in antisolar direction. The brightness gradient in the coma spike is reproducible
for a fast activity drop by two orders of magnitude within 3~h to 5~h after the
break-up time (which also defines the period of main dust production). Knight \& Battams (2014)
also postulated such a behavior in their preliminary analysis.
The dust expansion velocity is constrained by the opening angle of the coma spike
to extend not more than $\approx$0.5~km/s (which results in an opening angle of 12$\degree$,
while a dust expansion speed of 2~km/s would result in an angle of $\approx$17$\degree$).

%figure 4
\begin{figure}[t]
   \centering
   \includegraphics[width=9cm]{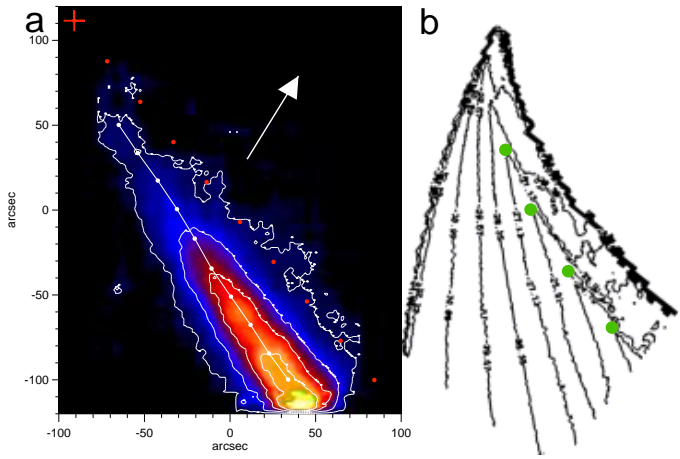}
      \caption{{\bf a:} Zoom of the last image in Fig.~2.
      The bi-sector of the tail is indicated by a white line.
      Predicted positions of the nucleus at mid-exposure for the last 10~minutes are shown as red dots (ending with a cross) and indicate the trajectory.
      The arrow points toward Sun center.
      The tail is offset from the trajectory and not parallel to it.
      {\bf b:} A model dust tail can reproduce the observation. The line of green dots marking
      the tip points of the higher-level contours resembles the tail axis,
      if we assume a violent dust emission burst $\approx$8.5 hours prior to perihelion transit
      followed by a sharp decrease of the dust production.
      }

\label{bisector}
\end{figure}

\section{Discussion}

We interpret the coma and tail appearance of the comet shortly before perihelion
transit as an indication for a rapid decrease of dust production,
which had ceased or dropped drastically by the time the comet entered the SUMER FOV, not showing any visible signature
of on-going activity. Although taken at a
different wavelength and at a much higher resolution, the appearance of the tail
in Fig.~4a can be understood as a continuation of the dimming and narrowing process
observed in LASCO C2 that finally leads to a complete disappearance of
nucleus activity. Interestingly, similar effects were reported by Bonev et al. (2002)
when they observed the disintegration of comet C/1999 S4 (LINEAR),
although this process occurred at a much larger distance from the Sun and took
several days to complete. The effects observed at comet ISON during its last hours
of activity seem to be similar, although much more accelerated.

Micron-sized and possible larger dust fills an arrow-shaped coma with a linear
slope of surface brightness in motion direction.
This coma pattern can be understood as a consequence of a rapidly decaying (few hours)
dust activity after a major nucleus break-up event that occurred about 8.5~h
(and definitely less than 12~h) before perihelion transit of the comet,
that is, after maximum brightness in the outburst on Nov 28.
It may thus represent the dusty aftermath of this nucleus splitting.
Most of the solids released during the outburst and before may have disappeared.
This might be due to evaporation of dust material in the very hot environment
close to the Sun where grains may have to experience temperatures of up to 2000~K.
%By the time of the SUMER observations the dust seen in the images within 0.014~AU from the Sun.
%Given the high temperature for the grains and considering the measured
%$\beta$ ratios of the dust, graphite is considered a more realistic material
%for the dust composition compared to silicatic basalt.

The low-level Ly-$\alpha$
flux of the dust coma is slightly asymmetric with respect to the main spike axis, with a
larger extension on the sunward side. This finding is not represented in the
isophote patterns of the dust coma simulations of the event as described above.
We note, however, that the enhanced low-level activity covers the trajectory
path of the nucleus through our FOV. Therefore one can assume
that additional dust exists on the sunward side of the arrow-shaped coma from
the ceasing activity of the comet. This dust is unrelated to the possibly
fatal break-up of the comet around Nov 28 at 10:15~UTC, but results from earlier
activity of the nucleus. Given its low intensity, the dust may
be subject to evaporation with significantly higher mass loss than at its release epoch.
%It is noted that the scenario introduced as explanation for the dust coma observed
%in the SUMER images requires the cometary nucleus position about {50\arcsec} to
%{100\arcsec} away along the trajectory path from the predicted one which is
%considerably more than the position uncertainty.

\section{Conclusion}

Unfortunately, the comet made it impossible to reach our prime scientific
goal, and no compositional information was obtained.
However, we managed to observe ISON's coma in Ly-$\alpha$ emission
only minutes before perihelion transit, when other instruments were paused because of
the close proximity of the solar disk. We interpret the signal as scattered light from the disk,
resembling a -8.5~h synchrone, and were able to model a tail with a spiky coma and brightness decrease in
motion direction, and with the nucleus not embedded in the dust cloud.
%Similar to the observed tail, the dust cloud is offset from the trajectory
%and has an asymmetric brightness profile.

Our dust dynamic model definitively requires a sharp fall-off of the dust production
hours before perihelion transit. Therefore our observations clearly indicate that the dust
production had already stopped before the comet entered our FOV and
that only the relicts of previous activity are seen.

The Ly-$\alpha$ radiance of the dust tail was four orders of magnitude below
the disk radiance, not too far from the detection limit of our instrument.
An enormous dynamical range is required for instruments that observe the comet
together with the disk. This may explain why no signature of the comet was seen in SDO-AIA data.

The SUMER Ly-$\alpha$ images of comet ISON, taken within an hour from perihelion transit,
are unique and show the cometary dust coma produced under very unusual circumstances.
To our knowledge, such observations have never been completed before. They
complement observations by other instruments, and their scientific value has to be seen
in concert with these other observations.

\begin{acknowledgements}

W.T. Thompson provided us with HPC coordinates as seen from SOHO. An exceptional effort
of D. Germerott and the SOHO flight operations team was required for this time-critical
observation. We received very constructive comments from an anonymous referee.
The help of all of them is greatly acknowledged. The SUMER and LASCO
projects are financially supported by DLR, CNES, NASA, and the ESA PRODEX
Programme (Swiss contribution). SUMER and LASCO are part of SOHO of ESA and NASA.

\end{acknowledgements}

\end{document}